# Doping-induced evolution of the intrinsic hump and dip energies dependent on the sample fabrication conditions in $Bi_2Sr_2CaCu_2O_{8+\delta}$

Tatsuya Honma

*Department of Physics, Asahikawa Medical University,*
*Asahikawa, Hokkaido 078-8510, Japan*
*honma@asahikawa-med.ac.jp*

## Abstract

In oxygen-doped $Bi_2Sr_2CaCu_2O_{8+\delta}$, the spectra, as observed by tunneling or photoemission spectroscopic measurements, depend on the sample fabrication conditions, such as the temperature and pressure for fabricating the junction or surface. This implies that the hump and dip energies extracted from the spectra depend on the sample fabrication conditions. When the samples were fabricated at 4.2 K and/or under ultra-high vacuum (UHV), the hump energy exhibited a new step-like doping dependence and the dip energy followed the upper pseudo-gap line. As the fabrication conditions deteriorated, the hump and dip energies reproduced the previous results, in that the dip energy was significantly dependent on the sample, whereas the hump energy exhibited a smooth doping dependence. It can be concluded that the observations for the samples fabricated at 4.2 K and/or under UHV reflect the intrinsic bulk properties, whereas those for the samples fabricated under deteriorated conditions reflect degraded surface properties.

*Keywords: high-temperature cuprate superconductor, electronic phase diagram, tunneling spectroscopy, photoemission, sample fabrication conditions.*

## 1. Introduction

One of the characteristic features of high-temperature cuprate superconductors (HTCS) is the so-called peak-dip-hump (PDH) structure, which can be observed in the energy distribution curve (EDC) by angle-resolved photoemission spectroscopy (ARPES).[1] The peak, dip, and hump structures appear away from the Fermi level ($E_F$) in the order. A similar PDH structure can also be observed in the spectrum below $E_F$ by superconductor-insulator-normal metal (SIN) tunneling spectroscopy.[2] However, the spectrum was not symmetrical with respect to $E_F$. The dip and hump structures above $E_F$ were sample dependent.[3-8] In contrast, the spectrum obtained by superconductor-insulator-superconductor (SIS) tunneling spectroscopy has a symmetrical PDH structure with respect to $E_F$.[6,9-12] Despite the different experimental probes and different resolutions, the peak and hump energies, defined as the visible peak positions of the peak and hump structures, followed the lower pseudo-gap (PG) and smooth hump lines in the unified electronic phase diagram (UEPD), respectively.[13] However, the dip energy, roughly defined as a visible bottom position between peak and hump structures, did not follow any line in the UEPD.

According to Giaever and Megerle,[14] the peak structure of the SIS tunneling spectrum can be well explained by a "peak-to-peak" tunneling process. Furthermore, both the symmetrical PDH structure in the SIS tunneling spectrum and the asymmetrical PDH structure in the SIN tunneling spectrum can be explained by the extended Giaever model (hereinafter referred to as the "single-hump model"), when it is assumed that the density of states (DOS) of HTCS has hump structure under peak structure below $E_F$. Then, the symmetrical PDH structure of the SIS tunneling can be explained in terms of two tunneling processes, "hump-to-peak" (Fig. 1(b)) and "peak-to-peak." This is consistent with the ARPES observations. From the SIS tunneling spectrum, the intrinsic hump and peak energies ($E_H^*$ and $E_P^*$) in the DOS were calculated as $|E_H^{SIS} - E_P^{SIS}/2|$ (Fig. 1(b)) and $|E_P^{SIS}|/2$, respectively. Here, $E_H^{SIS}$ and $E_P^{SIS}$ are the visible hump and peak energies in the SIS tunneling spectrum, respectively. The visible hump and peak energies ($E_H^{SIN}$ and $E_P^{SIN}$) in the SIN tunneling spectrum correspond to $E_H^*$ (Fig. 1(a)) and $E_P^*$, respectively. The visible hump and peak energies ($E_H^{PES}$ and $E_P^{PES}$) in the ARPES EDC also correspond to $E_H^*$ and $E_P^*$, respectively.

Purely oxygen-doped $Bi_2Sr_2CaCu_2O_{8+\delta}$ (OD-Bi2212) is a popular HTCS because it is easy to obtain a clean and shiny surface parallel to the $CuO_2$ plane (or *a*-*b* plane) by cleavage at room temperature (RT) and under ambient pressure. In 1995, Hancotte *et al*. reported that the tunneling spectrum through the SIN junction (SIN J) between OD-Bi2212 and a normal metal depends on whether the surface of OD-Bi2212 is cleaved at 4.2 K or RT.[15] Briefly, the SIN J fabricated at 4.2 K produced a symmetrical spectrum, whereas that fabricated at RT produced an asymmetrical spectrum. In 2010, Palczewski *et al*. reported that the ARPES EDC of OD-Bi2212 depended on the temperature and environmental pressure when the sample surface was cleaved.[16] Specifically, the surface cleaved at a low temperature under ultra-high vacuum (UHV) was stable until the temperature or pressure was increased. Accordingly, for OD-Bi2212, sample fabrication at 4.2 K and under UHV is effective for producing an ideal junction or surface. This finding, as suggested by Ref. 15, highlights the importance of the sample fabrication conditions, a factor not previously emphasized in electronic phase diagrams, such as the UEPD established in 2008.[13]

Hancotte *et al*. also observed another symmetrical hump structure outside the symmetrical PDH structure in the tunneling spectrum through an SIS break junction (BJ) fabricated at 4.2 K.[17] Unfortunately, the second symmetrical hump structure could not be explained by the above single-hump model. According to Mourachkine's interpretation,[18] the second symmetrical hump structure can be generated by a "hump-to-hump" tunneling process (Fig. 1(d)) when the DOS of HTCS has one pair of hump structures outside the paired peak structure (hereinafter referred to as the "twin-hump model"). Thus, $E_H^*$ can be calculated as $|E_{H2}^{SIS}|/2$ from the second hump structure (Fig. 1(*d*)), where $E_{H2}^{SIS}$ is the visible peak energy of the second hump structure. $E_H^*$ can also be calculated as $|E_H^{SIS}-E_P^{SIS}/2|$ (Fig. 1(c)).

Fortunately, the relationship of $E_H^* = |E_H^{SIS}-E_P^{SIS}/2|$ is independent of whether the hump-to-peak process is based on the single-hump model (Fig. 1(b)) or the twin-hump model (Fig. 1(c)). Accordingly, the data from tunneling and photoemission spectroscopic measurements in UEPD should be re-examined only in terms of "sample fabrication conditions."

In this study, the doping-induced evolution of $E_H^*$, as observed by tunneling or photoemission spectroscopic measurements in OD-Bi2212, was found to be significantly dependent on the sample fabrication conditions. When the samples were fabricated at 4.2 K and/or under UHV, $E_H^*$ exhibited a new step-like doping dependence. Furthermore, the intrinsic dip energy ($E_D^*$) corresponds to the upper PG line. However, as the fabrication conditions become poor, $E_H^*$ and $E_D^*$ reproduce previous results. Based on the observed effects of the sample fabrication conditions, it can be concluded that the observed results in the samples fabricated at 4.2 K and/or under UHV reflect an intrinsic bulk property, whereas the previous results reflect degraded surface properties. The dip energy can be considered as the boundary between the hump and peak contributions, although it is highly sensitive to sample fabrication conditions.

## 2. Method

The visible peak, dip, and hump energies for tunneling and photoemission spectroscopies were extracted from raw spectra[3-12,17-25] and raw EDC along the antinodal direction,[26-36] respectively, as reported in the literature. The sample fabrication conditions were determined from literature, as discussed below. The broad hump structure is sometimes flat. If it is difficult to determine the visible hump energy from the published figures, the center position is plotted as the hump energy. Then the flat region is shown as the error bars. The doped-hole concentration ($P_{pl}$) should be determined according to the hole scale based on the thermoelectric power at 290 K, as reported by Honma and Hor ($P_{pl}$-scale).[13,37] Unfortunately, the thermoelectric power was not accompanied by tunneling or ARPES data. Therefore, $P_{pl}$ was estimated by comparing the $T_c$ value in the literature with the half-dome-shaped $T_c$-curve in Ref. 13, because $T_c$ of the sample was provided along with its maximum value ($T_c^{max}$). In the case where $T_c^{max}$ was not provided, 93 K was used as $T_c^{max}$ for OD-Bi2212. The method for estimating $P_{pl}$ is provided in Ref. 38. Information on the half-dome-shaped $T_c$-curve for OD-Bi2212 is available in Supplementary Material.

## 3. Results

Figure 2 shows the $P_{pl}$-dependencies of $E_H^*$ and $E_P^*$ with respect to sample fabrication conditions. $E_H^*$ is significantly dependent on the sample fabrication conditions, whereas $E_P^*$ is nearly independent. $E_H^*$ and $E_P^*$ below $E_F$ are plotted in Figs. 2(a) - 2(c), whereas those above $E_F$ are plotted in Figs. 2(d) – 2(f). The data plotted in Fig. 2(a) are from the tunneling spectra obtained at 4.2 or 6-7 K through SIS BJs fabricated at 4.2 K[6,9-12,17-20] or approximately 10 K,[21] respectively. The details of these junctions are discussed later in this study. Note that $E_H^*$, which is calculated as $|E_H^{SIS}-E_P^{SIS}/2|$ or $|E_{H2}^{SIS}|/2$, does not follow the previous smooth doping dependence (dashed orange line) but follows a new step-like one (thick magenta line). The step-like hump line has five jumps at $P_{pl} \approx 0.17$, 0.19, 0.24, 0.27, and 0.28. Unfortunately, there was no available data at $0.19 < P_{pl} < 0.238$ ($0.91 < T_c/T_c^{max} < 1$ in the underdoped side). $E_P^*$ ($= |E_P^{SIS}|/2$) lies on the previous lower PG line within the error band,[13] although it appears to contain small structures. The result in Fig. 2(a) corresponds to that in Fig. 2(d). This is consistent with the observation of the symmetrical spectrum of the SIS tunneling. The plotted data in Fig. 2(b) are from the spectra through SIN Js fabricated at 4.2 K in He gas[6,7,15] or RT under UHV,[3,8,23-25] and the ARPES EDCs from surfaces cleaved under UHV.[26-30] $|E_H^{SIN}|$ and $|E_H^{PES}|$ in Fig. 2(b) exhibit the similar step-like doping dependence, although two jumps at $P_{pl} \approx$ 0.19 and 0.24 collapsed. Although $|E_P^{SIN}|$ and $|E_P^{PES}|$ follow the lower PG line, the distribution appears to expand toward the upper PG line. Gomes *et al.*[24] reported the distribution of peak energy, so-called gap distribution. The result is plotted in the Fig. 2(b) as the olive symbols for the mode energy with thick vertical bars for the distribution. The least upper bound of the distribution is around the collapsed step-like hump line, whereas the greatest lower bound is around the lower PG line. It seems that $E_P^*$ and $E_H^*$ get close to each other. These may be related to the partial overlap between peak and hump structures coming from the peak broadening induced by the poor fabrication condition. Whereas the results in Fig. 2(b) are consistent with those in Fig. 2(e), the plotted data in Fig. 2(e) are poor. This is coming from the asymmetrical spectra of the SIN tunneling and no information above $E_F$ of the ARPES.

The data in Fig. 2(c) are obtained from the spectra through SIN Js fabricated at RT[15] in a He atmosphere,[5] in flowing $N_2$[22] and in UHV (measurement for up to four weeks),[4] and the ARPES EDCs from the cleaved surfaces without special care.[31,32] $|E_H^{SIN}|$ and $|E_H^{PES}|$ reproduce the previous smooth hump line in the UEPD.[13] The spectra by Hancotte *et al.*[15] were unreproducible and hard to read the value of $E_H^{SIN}$ from their reported figure. Only the extracted value of $E_H^{SIN}$ follows the smooth hump line. Even though the SIN J was fabricated in UHV, it seems that a long period up to four weeks after the fabrication brings a change from the step-like hump to the smooth hump.[4] This may be related to the observation by Placzewski *et al.*[16]

The dip energy can also be classified based on sample fabrication conditions. In Figs. 3(a)-3(f), $E_D^*$ is plotted as a function of $P_{pl}$ using the same dataset as in Figs. 2(a)-2(f), respectively. Here, $E_D^{SIS}$, $E_D^{SIN}$, and $E_D^{PES}$ are the dip energies in the SIS tunneling spectrum, SIN tunneling spectrum, and ARPES EDC, respectively. Figure 3(a) shows the $P_{pl}$-dependence of $|E_D^{SIS}|/2$ at below $E_F$ from the SIS BJ fabricated at 4.2 and approximately 10 K. $|E_D^{SIS}|/2$ lies on the upper PG line at $P_{pl} < 0.20$, although it has a small upward bulge-like deviation from the upper PG line at $0.23 < P_{pl} <0.28$ (green thick line). Figure 3(b) shows the $P_{pl}$-dependence of $|E_D^{SIN}|$ and $|E_D^{PES}|$ at below $E_F$ from the SIN J fabricated at 4.2 K or RT under UHV and the surface cleaved under UHV, respectively. Although both $|E_D^{SIN}|$ and $|E_D^{PES}|$ roughly follow the upper PG line at $P_{pl} < 0.20$, they deviate upward from the upper PG line at $P_{pl} > 0.20$. The bulge-like deviation appears to grow toward the high-energy side, although some plotted data follow the green thick line obtained in Fig. 3(a). The plotted data distribute a range between the green thick line and the red thick line in the next Fig. 3(c). The sample history after the fabrication at the low temperature or in UHV may influence to the observation.[16] Figure 3(c) shows the $P_{pl}$-dependence of $|E_D^{SIN}|$ and $|E_D^{PES}|$ at below $E_F$ from the SIN J fabricated at RT or the cleaved surface without special care. The bulge-like deviation was the largest (red thick line).

Comparing the three figures of Figs. 3(a)-3(c), the bulge-like deviation appears to grow with deterioration of the sample fabrication conditions. The similar results are confirmed at above $E_F$ as shown in Figs. 3(d)-3(f). Thus, the dip energy is highly sensitive to the sample fabrication conditions.

Figure 4 shows the $P_{pl}$-dependence of the PG temperature ($T^*$) and other related temperatures/energies obtained using spectroscopic probes. In Fig. 4(a), all the plotted $T^*$ values[12,21,25,26,28,31,33-35,39] follow the upper PG line, although there is some scattering. The energy resolution of the ARPES data is below 8 meV[26,28,35] for the yellow symbols and over 15 meV[31,33,34] for the red symbols, and the energy resolution of the tunneling data (green symbols) is greater than 1 meV.[5] Accordingly, the $T^*$ values distributed between the upper and lower PG lines are considered to result from a relatively low energy resolution. Because the dip and peak energies follow the upper and lower PG lines, respectively, the peak structure appears at the upper PG line or PG temperature. In Fig. 4(b), the other reported characteristic temperatures follow either the smooth hump, upper PG, or lower PG lines. The pair formation temperature ($T_{c0}$) of intrinsic Josephson junctions (IJJs)[39] follows the lower PG line. Another PG temperature ($T_0$), referred to as the large PG temperature, was observed using angle-integrated photoemission spectroscopy (AIPES) by Takahashi *et al.*[28] $T_0$ is on the smooth hump line. $T^*$ reported by Matsuda *et al.*[22] also follows the smooth hump line. $T^*$ and pair formation temperature ($T_{pair}$) of ARPES by Kondo et al.[36] follow the smooth hump and upper PG lines, respectively. The temperature ($T_{100\%}$), where the sample surface is fully covered by the gapped area, by Gomes *et al.*[24] follows the lower PG line. The temperature ($T_{50\%}$), where a half of sample surface is covered by the gapped area,[24] tends to follow the upper PG line.

In Fig. 4(c), the other characteristic energies are plotted as a function of $P_{\rm pl}$. Howald *et al.*[23] observed a kink structure in the spectra of SIN tunneling. The similar kink structure was also observed in the spectra of the SIS tunneling.[7] A kink energy distributes around a half-dome-shaped $T_c$-curve or $3.5k_{\rm B}T_{\rm c}$-curve. Gomes *et al.*[24] reported a kink distribution based on the spectrum through the SIN J fabricated at RT. The mode energy lies on the $3.5k_{\rm B}T_{\rm c}$-curve. Note that the kink energy and the mode energy of the kink distribution follow the $3.5k_{\rm B}T_{\rm c}$-curve. The PG and superconducting gap ($\Delta_{\rm p}$ and $\Delta_{\rm S}$) by Ren *et al.*[39] are also plotted into Fig. 4(c). It seems that $\Delta_{\rm p}$ relatively follows the lower PG line, and $\Delta_{\rm S}$ also follows the lower PG line rather than the $3.5k_{\rm B}T_{\rm c}$-curve.

**4. Discussion**

The SIS BJs of the samples plotted in Figs. 2(a), 2(d), 3(a) and 3(d) were fabricated at 4.2 or approximately 10 K using the scanning tunneling microscope (STM) technique (or the point-contact technique; PCT) or the conventional BJ technique.

In the STM technique, the SIS BJ is formed by perforating the cleaved topmost layer by pushing the metal tip along the *c*-axis to the surface of the OD-Bi2212.[5,7] According to Ozyuzer's interpretation,[7] the insulator layer of the SIS BJ consists of adjacent Bi-O insulator layers, such as IJJs.[39] In the conventional BJ technique, the junction is formed by bending along the *a*-*b* plane at 4.2 K in a He atmosphere.[11,17] Neither of the tunneling currents passed through the topmost layer after cleavage. Thus, the SIS tunneling spectra through the SIS BJ fabricated at 4.2 or 10 K are relatively free from degradation caused by the cleaving process. Under these conditions, the probes can detect a pure virgin surface property, that is, the intrinsic bulk property. The SIN Js of the samples plotted in Figs. 2(b), 2(e), 3(b) and 3(e) were fabricated at 4.2 K in a He atmosphere or RT under UHV, and the surfaces of the samples for ARPES plotted in Figs. 2(b) and 3(b) were cleaved under UHV. Using the STM technique, an SIN J was formed by adjusting the pushing pressure of the tip against the cleaved surface.[7,40] A sufficiently high pressure formed the SIS BJ. However, when the pressure is either too low or too high, the junction does not function. A moderate pressure is required to form SIN J. Accordingly, the fabrication of SIN J may be more challenging than that of SIS BJ. Although cleavage under UHV can decrease the possibility of contamination by surface exposure,[16] this risk remains. This explains the collapsed step-like doping dependence of the hump energy in Fig. 2(b). The SIN Js formed by the STM technique in Figs. 2(c), 2(f), 3(c) and 3(f), except of one SIN J, were fabricated at RT in He atmosphere or flowing $N_2$. Although the SIN J by Renner *et al.*[4] was fabricated at RT under UHV, the junction was kept using spectra were measured for up to four weeks after fabrication. According to Ref. 15, the junctions fabricated at RT are unstable, except of the fabrication under UHV. According to Ref. 16, the relatively long period from the fabrication in UHV till the observation may cause the fresh surface to deteriorate. Thus, the tunneling currents pass through the degraded junction, and the photoelectrons are picked up from the surface degraded by cleavage. Under these conditions, the probes could detect the properties of the degraded surface. Therefore, the step-like hump behavior shown in Figs. 2(a) and 2(d) is a bulk property, whereas the smooth hump behavior shown in Figs. 2(c) and 2(f) reflects the properties of a degraded surface. In fact, as shown in Fig. 2, the distributions of hump and peak energies become wider in the order of Figs. 2(a), 2(b) and 2(c) (2(d), 2(e) and 2(f)).

As shown in Fig. 3, a bulge-like deviation in the dip energy was produced by the broadening of the hump and peak structures. Because the sharp peak had a higher intensity than the broad hump, the broadening effect of the peak was larger than that of the hump. Consequently, the bulge always grows toward the high-energy side because of broadening from degradation during sample fabrication. If the hump structure is away from the peak structures, the influence of the peak broadening can be eliminated. Therefore the dip energy in the underdoped side follows the upper PG energy line. When UEPD was established, the dip energy did not follow any lines in the UEPD. This issue originates from the sample fabrication issues. In Fig. 3, it seems that a lower energy resolution results in larger scattering. However, this influence did not directly contribute to the bulge-like deviation. Poor sample fabrication conditions prevented the identification of the intrinsic behavior of dip energy, particularly on the overdoped side. On the overdoped side, as the peak energy is closer to the hump energy, a partial overlap between the peak and hump structures becomes stronger. The dip energy depends on the degree of this overlap, resulting in a small bulge-like deviation at $0.23 < P_{\mathrm{pl}} < 0.28$ in Fig. 3(a). This deviation was enhanced by the deterioration of the sample fabrication conditions. Therefore, the dip energy following the upper PG line is an intrinsic property.

As shown in Fig. 4, $T^*$ values observed by the spectroscopic probes follow the upper PG line. The PG temperature was reported as a downward deviation from the $T$-linear dependence of the in-plane resistivity.[41] The resistive $T^*$ also follows the upper PG line.[13] Accordingly, the resistive downward deviation at $T^*$ can be related to the growth of the peak structure, that is, the enhancement of the DOS.

The three lines for the hump, upper PG, and lower PG energies in the previous UEPD were extracted from various characteristic energies/temperatures observed by different probes and groups.[13] Unfortunately, no single experimental probe detected all three energies of the hump, upper PG, and lower PG, although some probes detected only one or two energies. This study revealed that this problem stems from the sample fabrication issues. The hump, dip, and peak energies observed in the samples fabricated at 4.2 K and/or under UHV by tunneling and ARPES measurements correspond to the hump, upper PG, and lower PG energies in the UEPD, respectively.

$T^*$ and $T_{c0}$ of IJJ in Figs. 4(a) and 4(b) follow the upper and lower PG line, respectively. However, although $\Delta_p$ and $\Delta_S$ of IJJ in Fig. 4(c) were expected to follow the lower PG line and $3.5k_BT_c$-line, respectively, they seem to follow the lower PG line. This may be coming from the reported values of $\Delta_p$ and $\Delta_S$ based on the scaled spectrum.[39] The observed spectrum was scaled by spectrum at $T^*$ above which the spectrum becomes gapless.[39] Such scaling might bring a misestimation of the absolute energy value. However, the result in this study, except of some data in Fig. 4(c), is free from such issue, because the raw spectrum and EDC are analyzed.

The peak energy follows the lower PG line of the UEPD.[13] In the samples fabricated at 4.2 K, the intrinsic peak energy ($|E_P^{SIS}|/2$), which belongs to the lower PG, exhibited a minor structure, although it was buried in the error band of the lower PG line.[13] This structure appears to be related to the step-like hump behavior; however, it is challenging to check small structures because of resolution issues.

Further, the kink energy in Fig. 4(c) can be related to the $3.5k_BT_c$-curve or superconducting gap. There might be ambiguity issue in determining the kink energy. This is even more challenging to study.

The application of the $P_{pl}$-scale extracted a step-like doping dependence of $T_c$, which has three $T_c$ jumps at $P_{pl}$ = $P_4$, $2P_4$, and $3P_4$, in purely oxygen-doped $YBa_2Cu_3O_{6+\delta}$[42] and $La_2CuO_{4+\delta}$[43] from the oxygen-content (δ) dependency of $T_c$. Here, $P_4$ is equal to 1/16 based on 4 × 4 charge ordering. In the oxygen and cation co-doped $Bi_2Sr_{2-x}La_xCuO_{6+\delta}$, the insulator-superconductor (IS) transition appears at $P_{pl} \approx 0.11 \approx P_3$.[37] Here, $P_3$ is equal to 1/9 based on 3 × 3 charge ordering.[42] In OD-Bi2212, the $P_{pl}$-dependency of $E_H^*$ exhibits five jumps at $P_{pl} \approx 0.17$, 0.19, 0.24, 0.27, and 0.28. $P_{pl} \approx 0.17$, 0.19, 0.24, 0.27, and 0.28 are corresponding to 1/6, 3/16, 17/72, 77/288, and 9/32, respectively. Based on the hierarchical behavior, the $P_{pl}$ values of 1/6, 3/16, 17/72, 77/288, and 9/32 can be represented as $(P_3+2P_3)/2 = 1.5P_3$, $3P_4$, $(2P_3+4P_4)/2 = P_3+2P_4$, $(2P_3+5P_4)/2$, and $(4P_4+5P_4)/2 = 4.5P_4$, respectively. In the half-dome-shaped $T_c$-curve of OD-Bi2212 (red broken line in Figs. 2 and 3),[13] IS transition, $T_c^{max}$, and $T_c$-disappearance occur at $P_{pl} \approx 0.09$, 0.24, and 0.31, corresponding to 3/32, 17/72, and 5/16, respectively. The values of $P_{pl}$ = 3/32, 17/72, and 5/16 can be represented by $(P_4+2P_4)/2 = 1.5P_4$, $(2P_3+4P_4)/2$, and $5P_4$, respectively. Thus, the characteristic features of OD-Bi2212 appear in a hierarchical order based on $P_4$ and the mixing level of the nearest $P_4$ or $P_3$. Although OD-Bi2212 is a purely oxygen-doped HTCS, it has an incommensurate superlattice structure along the *b*-axis.[44] The superlattice structure may influence this hierarchical order.

**5. Conclusion**

The doping-induced evolution of $E_H^*$ and $E_D^*$, as extracted from the tunneling spectra or ARPES EDC of OD-Bi2212, was studied in terms of sample fabrication conditions. When the samples were fabricated at 4.2 K and/or under UHV conditions, $E_H^*$ exhibited step-like doping dependence and $E_D^*$ followed the upper PG line. However, as the fabrication conditions deteriorated, the doping dependence of $E_H^*$ and $E_D^*$ reproduced the previous behavior observed in the samples fabricated at RT in exposing gas. Therefore, it can be concluded that $E_H^*$ and $E_D^*$ of the samples fabricated at 4.2 K and/or under UHV conditions reflect the bulk properties, whereas those of the samples fabricated under poor conditions reflect degraded surface properties. Unfortunately, the bulk property cannot be resolved against the intrinsic peak energy and superconducting gap energy because of the resolution limitations of experimental probes. Therefore, a special analytical technique is required to identify intrinsic features from extensive and reliable data on OD-Bi2212. This study is currently in progress.

**Acknowledgements**

The author appreciates the collaboration with Dr. Pei-Herng Hor (Department of Physics and Texas CTR for Superconductivity at the University of Houston) for over 20 years. Unfortunately, he passed away in the early spring of 2021. The author declares no conflicts of interest. This study did not receive any specific grants from funding agencies in the public, commercial, or not-for-profit sector. No new data in the study or experimental data were obtained from the published papers.

**Figure captions**

**Figure 1.** (Color online) Schematic of the SIN or SIS tunneling related to the hump energy of the HTCS at 0 K (based on Refs. 14 and 18). (a) SIN tunneling with $V_{\mathrm{Bias}} = E_{\mathrm{H}}^{*}/e$ (single-hump). (b) SIS tunneling with $V_{\mathrm{Bias}} = (E_{\mathrm{P}}^{*} + E_{\mathrm{H}}^{*})/e$ (single-hump). (c) SIS tunneling with $V_{\mathrm{Bias}} = (E_{\mathrm{P}}^{*} + E_{\mathrm{H}}^{*})/e$ (twin-hump). (d) SIS tunneling with $V_{\mathrm{Bias}} = 2E_{\mathrm{H}}^{*}/\mathrm{e}$ (twin-hump). $V_{\mathrm{Bias}}$ and $e$ are the bias voltage and electron charge, respectively. The thick arrows show the main tunneling process.

**Figure 2.** (Color online) $P_{pl}$-dependencies of $E_H^*$ and $E_P^*$. (a) and (d) Based on the spectra through the SIS BJs fabricated at 4.2[6,9-12,17-20] or approximately 10 K.[21] (b) and (e) Based on the spectra through the SIN Js fabricated at 4.2 K[6,7,15] or at RT under UHV,[3,8,23-25] and the EDCs from the surfaces cleaved under UHV.[26-30] (c) and (f) Based on the spectra through the SIN J fabricated at RT[15] in He atmosphere,[5] in flowing $N_2$[22] and in UHV (up to 4 weeks),[4] and the EDC from the surface cleaved at RT.[31,32] Although the SIN J of Ref. 4 was fabricated in UHV, the junction was used for up to four weeks. The olive symbols and bold lines show the mode energy and distribution range of the peak energy (gap distribution) in Ref. 24, respectively. All orange and green symbols show the hump and peak energies, respectively. The half-black symbols in Figs. 2(a) and 2(d) show $|E_{H2}^{SIS}|/2$. The energy resolution of the tunneling[5] and laser-based ARPES[26,27] is about 1 meV. The energy resolution of ARPES shown as half-black symbol is below 8 meV[28-30] in Fig. 2(b) and over 15 meV[31,32] in Fig. 2(c), respectively. The five vertical broken lines show $P_{pl}$ = 1/6, 3/16, 17/72, 77/288, and 9/32. The tunneling data are observed at 4.2 K. The other temperatures, including for ARPES, are noted in parentheses of each caption. In Figs. 2(b) and 2(c), almost all SIN Js are fabricated by STM tech. or PCT. Only the vacuum junction (VJ) is noted in the parentheses.

**Figure 3.** (Color online) $P_{\mathrm{pl}}$-dependence of $E_{\mathrm{D}}^{*}$. (*a*) Based on the same dataset in Fig. 2(a). (b) Based on the same dataset in Fig. 2(*b*). (*c*) Based on the same dataset in Fig. 2(c). (d) Based on the same dataset in Fig. 2(d). (e) Based on the same dataset in Fig. 2(e). (f) Based on the same dataset in Fig. 2(f). The green and red thick lines are guide to the eyes.

**Figure 4.** (Color online) $P_{\mathrm{pl}}$-dependence of $T^*$ and the other related temperature. (a) $P_{\mathrm{pl}}$-dependence of $T^*$. $T^*$ values are from SIS tunneling,[12,20] SIN tunneling,[25] laser-based ARPES,[26] ARPES,[28,31,33-35] and IJJs.[39] The energy resolution of ARPES is below 8 meV for the yellow symbols[26,28,30] and over 15 meV for the red symbols.[31,33,34] The energy resolution of the tunneling[5] and laser-based ARPES[26] is 1 meV. (b) $P_{\mathrm{pl}}$-dependence of the other related temperature. $T_{\mathrm{c0}}$ and $T_0$ are from IJJs[21] and AIPES,[28] respectively. $T^*$ by tunneling is coming from Ref. 22. $T^*$ and $T_{\mathrm{pair}}$ by ARPES are coming from Ref. 36. $T_{50\%}$ and $T_{100\%}$ are from STM.[24] (c) $P_{\mathrm{pl}}$-dependence of the other characteristic energy.[7,21,23-25] The temperature ($T$) scale and energy ($E$) scale are related by $E = 3.5k_{\mathrm{B}}T$,[13] where $k_{\mathrm{B}}$ is the Boltzmann constant. The open symbols show the lower bounds (LB) for $T^*$.[31,34] The energy resolutions of ARPES[36] and AIPES[28] are 10 and 7 meV, respectively.

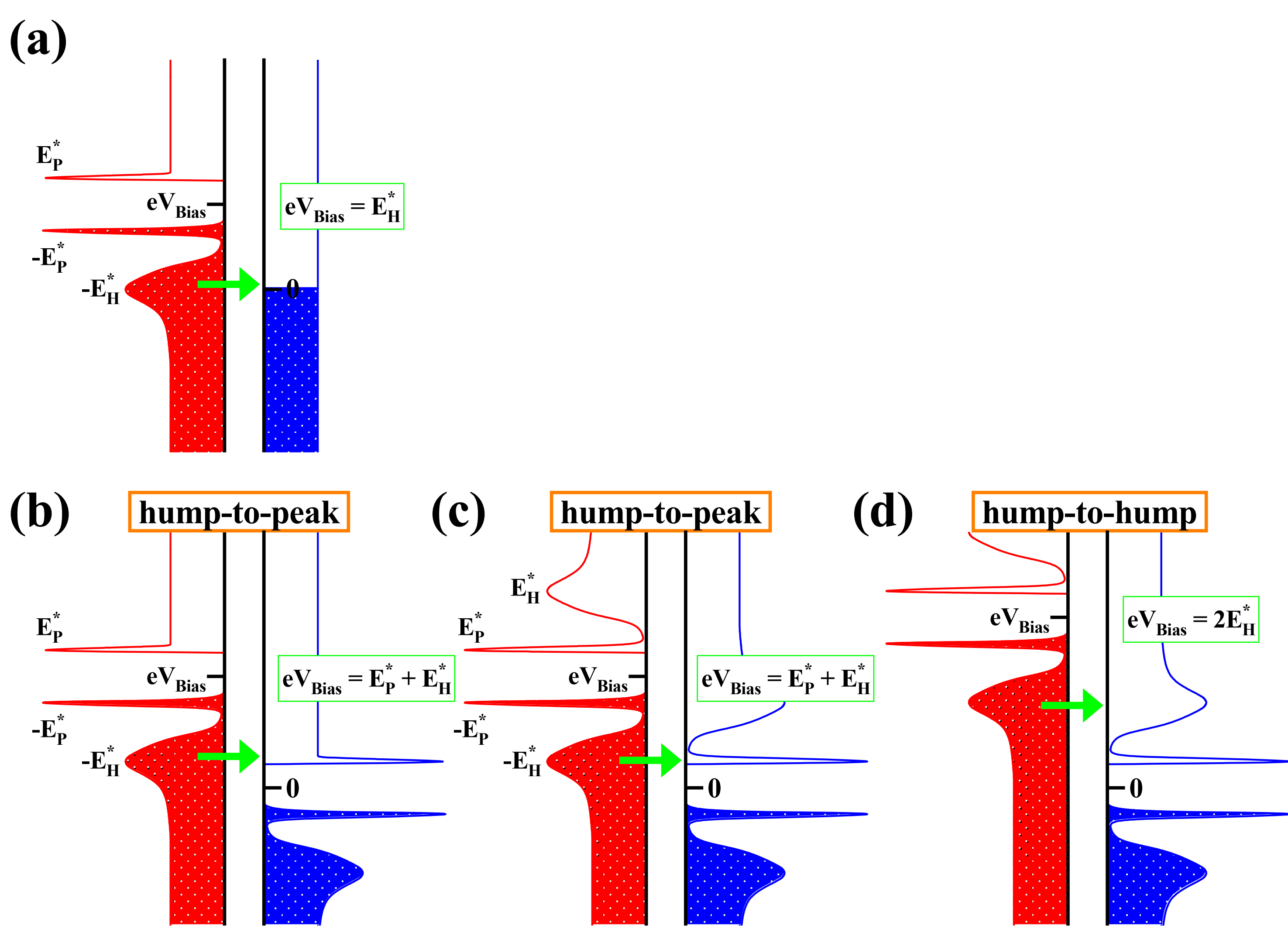

**Figure 1 Honma**

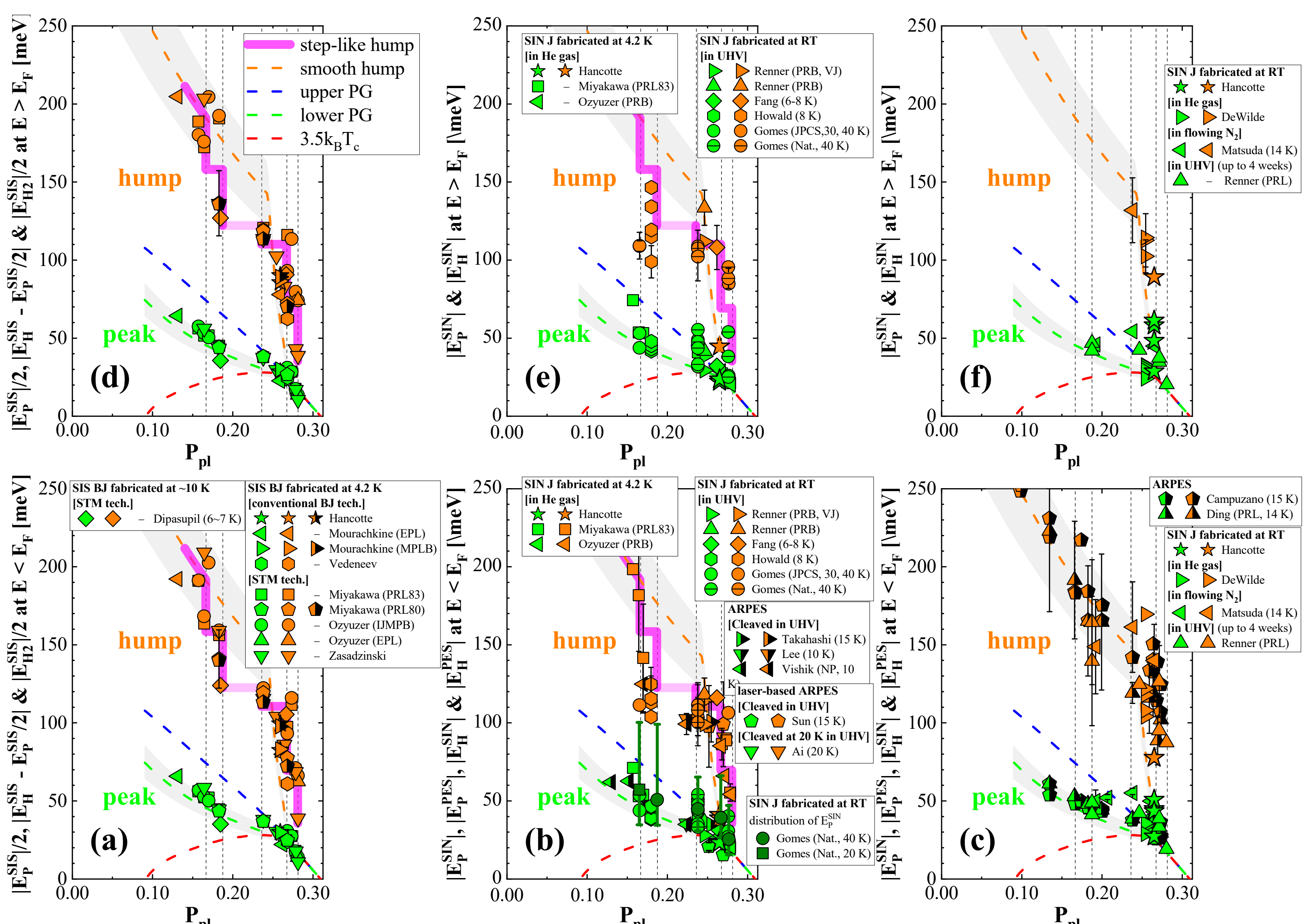

**Figure 2 Honma**

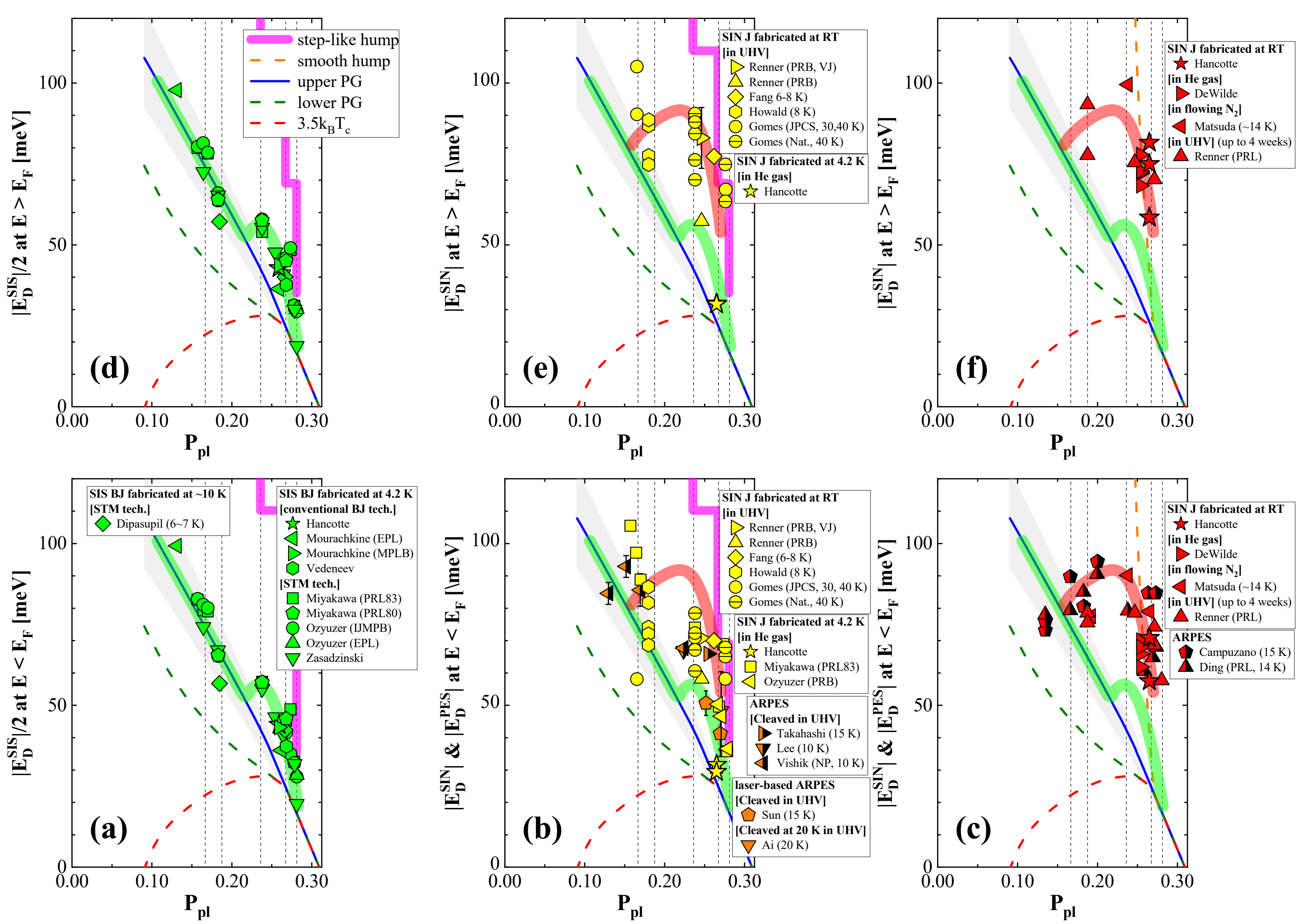

**Figure 3 Honma**

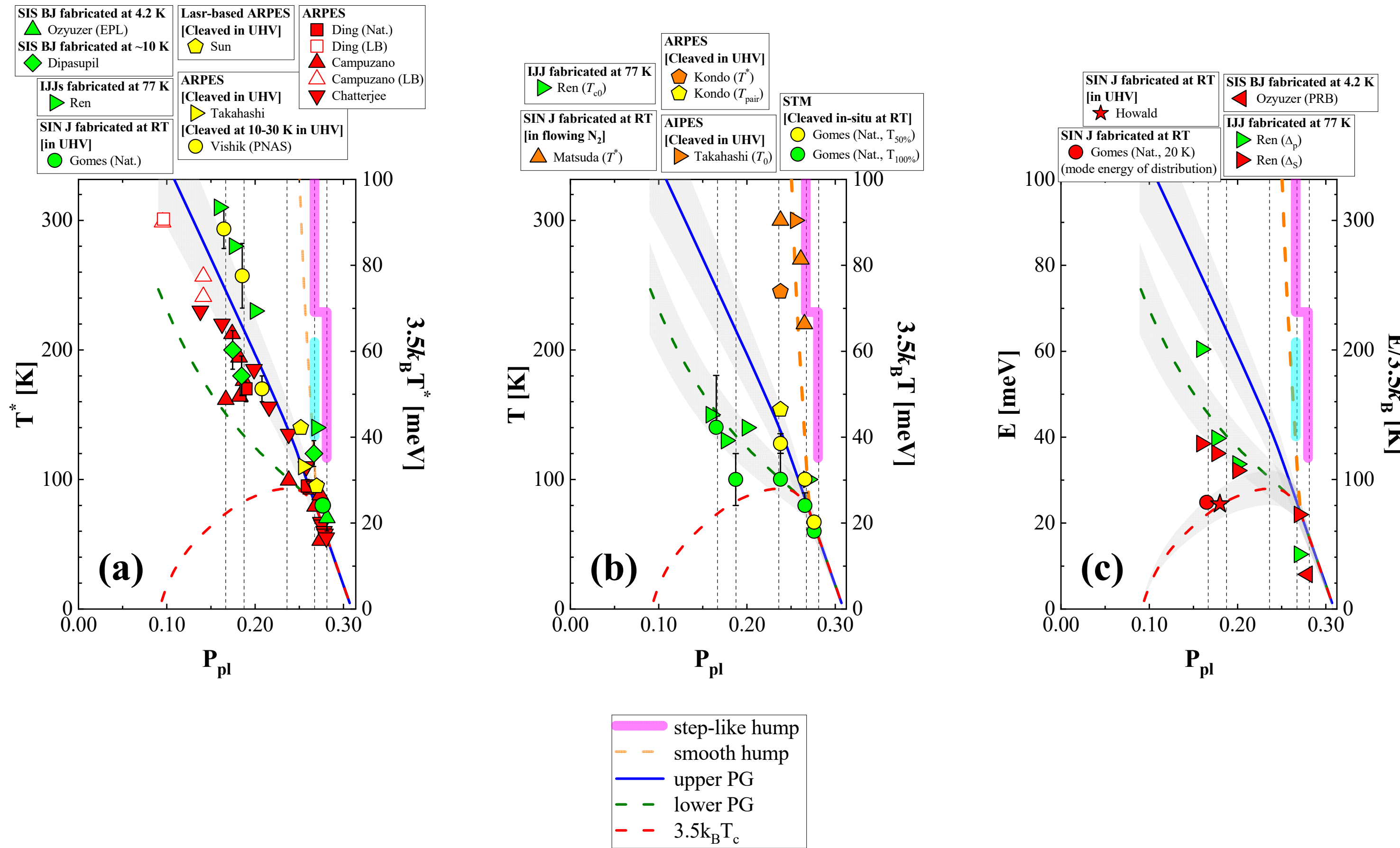

# Figure 4 Honma

# Supplementary Material for the doping-induced evolution of the intrinsic hump and dip energies dependent on the sample fabrication conditions in $Bi_2Sr_2CaCu_2O_{8+\delta}$

Tatsuya Honma

Department of Physics, Asahikawa Medical University, Asahikawa 078-8510, Japan

## Determination of $P_{pl}$

A detailed description of $P_{pl}$ was published by Honma and Hor.[13,37,38] The first and most reliable method for extracting $P_{pl}$ is to determine the value of $P_{pl}$ from the thermoelectric power at 290 K (S290) using $P_{pl}$-scale.[13,37] In the second method, $P_{pl}$ is determined from $T_c$ by comparison with a universal half-dome-shaped $T_c$-curve.[13] The universal half-dome-shaped $T_c$-curve was based on $P_{pl}$ determined from S290. The $T_c$-curve of OD-Bi2212 is shown in Fig. S1. The values of these data points are listed in Table S1. Unfortunately, in the present analysis, the value of S(290) is not available in the literature. Therefore, the determination of $P_{pl}$ was based on a second method. The $P_{pl}$ value was estimated from the $T_c$-curve shown in Fig. S1, after $T_c$ of the samples was scaled by $T_c^{max}$.

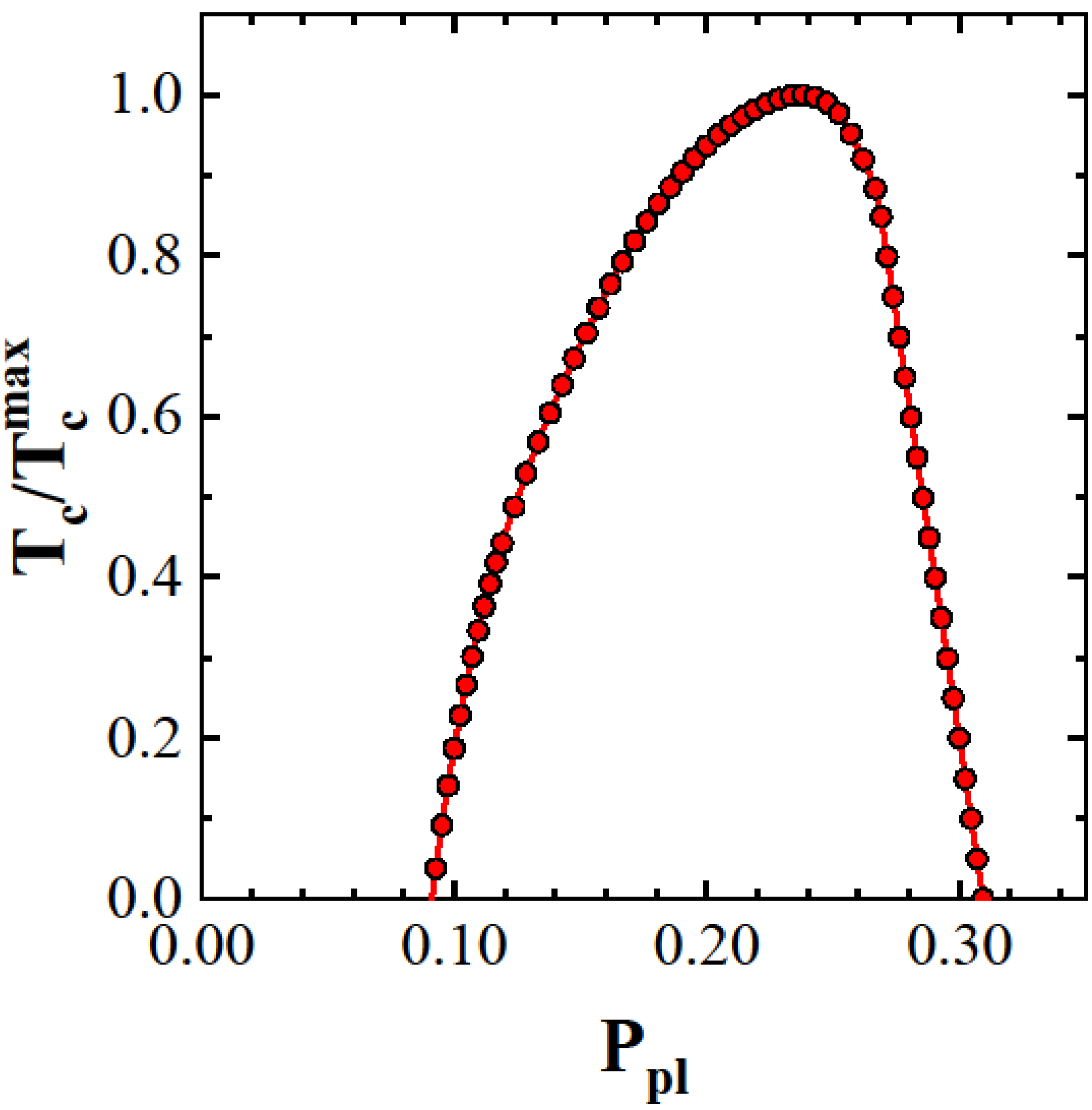

**Fig. S1.** (Color online) A half dome-shaped $T_c$-curve for $Bi_2Sr_2CaCu_2O_{8+\delta}$.[13] The data for the red circles are listed in the table S1.

| $T_c/T_c^{max}$ | $P_{pl}$ |
|---|---|
| 0.0375 | 0.0928 |
| 0.0918 | 0.0952 |
| 0.1413 | 0.0976 |
| 0.1867 | 0.1000 |
| 0.2282 | 0.1023 |
| 0.2663 | 0.1047 |
| 0.3014 | 0.1071 |
| 0.3339 | 0.1095 |
| 0.3639 | 0.1119 |
| 0.3919 | 0.1142 |
| 0.4181 | 0.1166 |
| 0.4428 | 0.1190 |
| 0.4882 | 0.1238 |
| 0.5296 | 0.1285 |
| 0.5681 | 0.1333 |
| 0.6044 | 0.1380 |
| 0.6390 | 0.1428 |
| 0.6723 | 0.1476 |
| 0.7043 | 0.1523 |
| 0.7350 | 0.1571 |

| $T_c/T_c^{max}$ | $P_{pl}$ |
|---|---|
| 0.7643 | 0.1618 |
| 0.7921 | 0.1666 |
| 0.8182 | 0.1714 |
| 0.8426 | 0.1761 |
| 0.8651 | 0.1809 |
| 0.8856 | 0.1856 |
| 0.9043 | 0.1904 |
| 0.9212 | 0.1952 |
| 0.9364 | 0.1999 |
| 0.9499 | 0.2047 |
| 0.9620 | 0.2094 |
| 0.9726 | 0.2142 |
| 0.9818 | 0.2190 |
| 0.9895 | 0.2237 |
| 0.9953 | 0.2285 |
| 0.9990 | 0.2332 |
| 1.0000 | 0.2380 |
| 0.9974 | 0.2428 |
| 0.9903 | 0.2475 |
| 0.9774 | 0.2523 |

| $T_c/T_c^{max}$ | $P_{pl}$ |
|---|---|
| 0.9513 | 0.2570 |
| 0.9196 | 0.2618 |
| 0.8837 | 0.2666 |
| 0.8485 | 0.2689 |
| 0.7986 | 0.2713 |
| 0.7486 | 0.2737 |
| 0.6987 | 0.2761 |
| 0.6488 | 0.2785 |
| 0.5989 | 0.2808 |
| 0.5490 | 0.2832 |
| 0.4991 | 0.2856 |
| 0.4492 | 0.2880 |
| 0.3993 | 0.2904 |
| 0.3494 | 0.2927 |
| 0.2995 | 0.2951 |
| 0.2495 | 0.2975 |
| 0.1996 | 0.2999 |
| 0.1497 | 0.3023 |
| 0.0998 | 0.3046 |
| 0.0499 | 0.3070 |

**Table S1** Scaled $T_c$ ($T_c/T_c^{max}$) and the corresponding value of $P_{pl}$ for the red circles plotted in Fig. S1.